\documentclass[twocolumn,floatfix,superscriptaddress,amsmath,showkeys,aps,prb,eqsecnum]{revtex4-1}
\usepackage{bm}
\usepackage{xcolor}
\usepackage{hyperref}
\usepackage{makeidx}
\usepackage{amsmath}
\usepackage{graphicx}
\usepackage{dcolumn}
\usepackage{amsfonts}
\usepackage{amssymb}
\usepackage{color}

\begin{document}
\title{Influence of the symmetry of the hybridization on  the critical temperature of multi-band superconductors}

\author{Daniel Reyes}
\email{daniel@cbpf.br}
\affiliation{Instituto Militar de Engenharia - Pra\c{c}a General Tib\'{u}rcio, 80, 22290-270, Praia Vermelha, Rio de Janeiro, Brazil}
\author{Nei Lopes}
\affiliation{Centro Brasileiro de Pesquisas F\'{\i}sicas, Rua Dr. Xavier Sigaud 150, Urca, 22290-180, Rio de Janeiro , Brazil}
\author{Mucio A. Continentino}
\affiliation{Centro Brasileiro de Pesquisas F\'{\i}sicas, Rua Dr. Xavier Sigaud 150, Urca, 22290-180, Rio de Janeiro , Brazil}
\author{Christopher Thomas}
\affiliation{Departamento de F\'isica, Universidade Federal Rural do Rio de Janeiro, 23897-000, Serop\'edica, Rio de Janeiro, Brazil}

\date{\today}

\begin{abstract}
In this work we study a two-band model of a superconductor in a square lattice. One band is narrow in energy and includes local Coulomb correlations between its quasi-particles. Pairing occurs in this band due to  nearest neighbor attractive interactions. Extended $s$-wave, as well as $d$-wave symmetries of the superconducting order parameter are considered. The correlated electrons hybridize with those in another, wide  conduction band through a $k$-dependent mixing, with even or odd parity depending on the nature of the orbitals.  The many-body problem is treated within a slave-boson approach that has proved adequate to deal with the strong electronic correlations that are assumed here.  Since applied pressure changes mostly the ratio between hybridization and band-widths, we can use this ratio as a control parameter to obtain the phase diagrams of the model.  We find that for a wide range of parameters, the critical temperature increases as a function of hybridization (pressure), with a region of  first-order transitions. When frustration is introduced it gives rise to a stable superconducting phase. We find that superconductivity can be suppressed for specific values of band-filling due to the Coulomb repulsion. We show how pressure, composition and strength of correlations  affect the superconductivity for different symmetries of the order parameter and the hybridization.
\end{abstract}

\maketitle

\section{Introduction}
Many superconducting materials of current interest  are multi-band systems, with electrons  from different atomic orbitals coexisting at a common Fermi surface. This is the case of heavy fermions and  high T$_c$ superconductors, either based on $Cu$ or $Fe$. Therefore, to take into account the multi-band character seems to be essential to understand the physical properties of these superconductors. 

In the materials we are interested, we can distinguish two different types of electronic quasi-particles. One is associated with nearly localized electrons in a narrow band, and another  consists of  conduction electrons in a wide band~\cite{Micnas,Robas87,Hewson,Steglich}.  The admixture between these distinct quasi-particles is responsible for many of the properties of multi-band systems, such as, their magnetism and response functions.

A model to describe these materials must consider these features and take into account the local Coulomb repulsion  among the electrons in the narrow $d$ (cuprates, pnictides) or $f$ (heavy fermions) bands. The most favorable conditions for the appearance of superconductivity involve an attractive interaction that pairs quasi-particles of the narrow band in neighboring sites avoiding in this way the strong on-site Coulomb repulsion. This type of pairing  entails different symmetries of the superconducting order parameter. Experimentally, it is well known that in the case of the cuprates, they adopt a $d$-wave symmetry and on-site pairing vanishes.

An interesting experimental fact observed  in multi-band superconductors is the existence of a  quantum phase transition associated with a superconducting quantum critical point (SQCP)~\cite{Ramos,Jin1,Jin2}. Varying an external control parameter, such as doping or pressure, these systems can be driven to a non-superconducting metallic state  or even to an insulator~\cite{Daniel16}. Notice that the ratio between hybridization and bandwidth, which  depends on the overlap of different wave functions is sensitive to these external parameters. Consequently, theoretical phase diagrams~\cite{Daniel16,Araujo0,Peres2,Oliveira,Koga1,Koga2,Daniel15}  obtained as a function of this ratio have a direct resemblance to those obtained experimentally when pressure or doping is varied~\cite{Ramos,Jin1,Jin2,Neto,Bauer1,Bauer2}. It turns out that in realistic cases, the parity of the hybridization is very important.  In multi-band systems we have in general to consider the mixing of $s$-$p$, $s$-$f$, $p$-$d$, $p$-$f$ and $d$-$f$ orbitals, which hybridize with different parities~\cite{deus,filipe}. 

The motivation of this work is to investigate how quantities, such as, the critical temperature vary as a function of  the different parameters of the model. Specifically, we consider the  intensity and symmetry of the hybridization,  the occupation of the bands and the strength of the local repulsion. 

We shall apply the slave-boson (SB) mean-field approach to solve the many-body problem. This is a well-known technique to deal with the type and magnitude of the electronic correlation that we are interested. We perform a self-consistent numerical solution of a set of coupled equations to obtain  finite temperature phase diagrams. 
These present  regions of metastability, i.e., of first-order transitions that mainly occur whenever the critical temperature increases as a functions of hybridization.  We show that if frustration effects are taken into account the superconducting phase can be stabilized in these regions. We  find that Coulomb interactions can  suppress superconductivity, for specific values of band-filling. 

We point out that our approach does not aim to describe any specific material, but to provide   a general guide to expectations of how different control parameters  affect the properties of multi-band superconductors when different symmetries of the hybridization and order parameter are considered.

The paper is organized as follows: in section \ref{sec2} we present the model, with its main features required to describe a multi-band superconductor with a narrow band of strongly correlated electrons. We also introduce in this section the SB formalism used to treat the many-body problem. In section \ref{sec3} we show and discuss our results for both, zero and finite temperature phase diagrams as a function of hybridization, band-filling and Coulomb repulsion. Finally, in section \ref{sec4} we point out the main results.

\section{The model \label{sec2}}

We consider a two-dimensional (2D), two-band lattice model with inter-site attractive interactions and  local Coulomb repulsion between electrons in the narrow band, which we refer generically as $f$-{\it electrons}.  These can be either $d$-electrons as for the $Cu$ or $Fe$ superconductors, or $f$-electrons as for the actinides and rare-earth heavy fermions. This narrow band hybridizes with a conduction band of $c$-{\it electrons} through a $k$-dependent hybridization that can have different symmetries.

The Hamiltonian of the model is given by~\cite{Daniel16},
\begin{eqnarray}
\label{hamilt}
\mathcal{H}&=&\sum\limits_{k,\sigma}\epsilon_{k}^{c}c_{k,\sigma}^{\dagger}c_{k,\sigma}+\sum\limits_{k,\sigma}
\epsilon_{k}^{f}f_{k,\sigma}^{\dagger}f_{k,\sigma}\nonumber \\
&+&\sum_{k,\sigma}(V_k c_{k,\sigma}^{\dagger}f_{k,\sigma}+h.c.)
+U\sum_{i}f_{i,\uparrow}^{\dagger}f_{i,\uparrow}f_{i,\downarrow}^{\dagger}f_{i,\downarrow}\nonumber \\
&+&\frac{1}{2}\sum_{\langle i j\rangle,\sigma}J_{ij}f_{j,\sigma}^{\dagger}f_{j,-\sigma}^{\dagger}f_{i,-\sigma}f_{i,\sigma}, \
\end{eqnarray}
where  $c_{k,\sigma}^{\dagger}$ ($c_{k,\sigma}$) and $f_{k,\sigma}^{\dagger}$ ($f_{k,\sigma}$) are creation (annihilation) operators related to conduction and $f$-electrons with spin $\sigma$ in the wide uncorrelated band, and in  the narrow band, respectively. 

These bands are described by the dispersion relations, $\epsilon_{k}^{c}$ and $\epsilon_{k}^{f}$, for $c$ and $f$-electrons in an obvious notation. Since we do not consider magnetic solutions these dispersions are independent of the spin $\sigma$.  $U$ is the on-site repulsive interaction
($U>0$) among the $f$-electrons. The two types of electrons are hybridized, with a $k$-dependent matrix element $V_k$~\cite{Daniel15}.  The last term describes an effective attraction between $f$-electrons in neighboring sites  {\bf($J_{ij} < 0$)} and is responsible for superconductivity~\cite{Sacramento}. It takes into account, experimental results for the specific heat in multi-band superconductors that show unequi\-vocally that $f$-electrons are involved in the pairing~\cite{Coleman3}. 
Notice that this term also describes antiferromagnetic (AF), $xy$-type  exchange interactions between these electrons, such that  magnetic and superconducting ground states are in competition. In this work we are only interested in the latter. We have also neglected in this interaction an Ising term that when decoupled in the superconducting channel leads to $p$-wave pairing that is not considered here. 

It is worth to point out that all terms in Eq.~(\ref{hamilt}) can vary substantially from one specific class of systems to another. In particular, the interactions $J_{ij}$ should be small for the case of rare-earth heavy fermions due to the localization of the $f$-orbitals in these systems. We have glossed over inter-band attractive interactions among the $c$ and $f$-electrons and  intra-band attractive interactions between $c$-electrons since they do not affect the superconducting order parameter in a drastic way~\cite{Sacramento}. Also inter-band pair hopping~\cite{Yosida} that arises in second order in the hybridization when applying a Schrieffer-Wolff transformation for the Anderson lattice model is ruled out, since it gives rise not only to ground states with finite $q$-pairing states but also to anisotropic $s$-wave superconductivity~\cite{Yamamoto}.

The  Hamiltonian,  Eq.~(\ref{hamilt}) represents a model that considers the basic features of a multi-band superconductor with a narrow band of strongly correlated electrons. It reflects a difficult many-body problem and has been treated using different approximations.  Se\-veral approaches have been used depending on the different aspects of the problem that one wants to emphasize. This includes, competition between different ground states~\cite{Becca}, superconducting properties at zero or finite temperature~\cite{Sacramento1,Spalek4}, $c$-$f$-pairing~\cite{Yamamoto} or the nature of the phase diagram as a function of the occupation of the bands. 

In this work, we approach this complicated many-body problem  using the mean-field slave boson formalism~\cite{Coleman1,Coleman2,Kotliar},  which  has been shown suitable for studying  coexistence between superconductivity and magnetism~\cite{Sacramento1}, crossover from BCS-type to local pairing~\cite{Robas}, magnetic instabilities~\cite{Wolfle,Dorin1,Aparicio}, and the effect of  infinite~\cite{Oliveira,Araujo,Peres} and finite~\cite{Daniel16,Sacramento,Dorin2} Coulomb repulsion in narrow bands.  It has also been shown to be in remarkable agreement with more elaborated numerical Monte Carlo results over a wide range of interactions and particle densities~\cite{Lilly}.

Considering a finite on-site Coulomb repulsion $U$ in the SB formalism, each lattice site one can have four physical states. The empty state $\mid0\rangle$,
the states where there is one $f$-electron with a given spin
$\mid\uparrow \rangle$ and $\mid \downarrow \rangle$ and the doubly occupied configuration  $\mid\uparrow \downarrow\rangle$, such that  the total number of $f$-electrons per site, $n_f$ can be  $n_f=n_{i\uparrow}^{f} +n_{i \downarrow}^{f}=0$,  $1$  or  $2$. In order to describe all these states that $f$-electrons can occupy, it is introduced four bosons $e$, $d$, $p_\uparrow$, and $p_\downarrow$, where $e$, $d$ are associated with empty and doubly occupied sites, respectively, and the boson $p_{\uparrow}$ ($p_{\downarrow}$) with a singly occupied site with spin
component $\uparrow$ ($\downarrow$)~\cite{Kotliar}.

For the purpose of establishing a one-to-one correspondence between the original Fock space and the enlarged one that also contains the bosonic states, the constraints $e_i^{\dagger}e_i+p_{i,\uparrow}^{\dagger}p_{i,\uparrow}+p_{i,\downarrow}^{\dagger}p_{i,\downarrow}+d_i^{\dagger}
d_{i}=1$ for the completeness of the bosonic operators, and $f_{i \sigma}^{\dagger}f_{i \sigma}=p_{i \sigma}^{\dagger}p_{i \sigma}+d_{i}^{\dagger}d_{i}$, for the local particle (boson+fermion) conservation at the $f$ sites,  must be satisfied. Such constraints are imposed in each site by the Lagrange multipliers $\lambda_i$ and $\alpha_{i,\sigma}$, respectively. 

In the physical sub-space, the operators $f_{i,\sigma}$ are mapped, such that, $f_{i,\sigma} \rightarrow f_{i,\sigma}Z_{i,\sigma}$ where 
$Z_{i,\sigma}=\frac{(e_{i}^{\dagger}p_{i,-\sigma}+p_{i,\sigma}^{\dagger}d_{i})}{\sqrt{(1-d_{i}^{\dagger}d_{i}-p_{i,\sigma}^{\dagger}p_{i,\sigma})(1-e_{i}^{\dagger}e_{i}-p_{i,-\sigma}^{\dagger}p_{i,-\sigma})}}$ and where the square root term  ensures that the mapping becomes trivial at the mean-field level in the non-interacting limit ($U\rightarrow 0$). The usual procedure consists in taking a mean-field approach where we assume the slave bosons to be condensed~\cite{Daniel16,Sacramento,Dorin1}. Then all bosons operators are replaced by their expectation values as,
$Z=\langle Z_{i,\sigma}^{\dagger}\rangle=\langle Z_{i,\sigma}\rangle=Z_{\sigma}$,
$e=\langle e_i\rangle=\langle e_i^{\dagger}\rangle$,
$p_{\sigma}=\langle p_{i,\sigma}\rangle=\langle p_{i,\sigma}^{\dagger}\rangle$, and
$d=\langle d_i\rangle=\langle d_i^{\dagger}\rangle$.
Due to translation invariance these expectation values take the same value on all sites.

Using this approximation the Hamiltonian  Eq.~(\ref{hamilt}) can be written as,
\begin{eqnarray}\label{effmodel}
\mathcal{H}_{eff}&=&\sum_{k,\sigma}(\epsilon_{k}^{c}-\mu)c_{k,\sigma}^{\dagger}c_{k,\sigma}+
\sum_{k,\sigma}(\tilde{\epsilon}_{k}^{f}-\mu)f_{k,\sigma}^{\dagger}f_{k,\sigma}\nonumber\\
&+&\sum_{k,\sigma} Z V_k (c_{k,\sigma}^{\dagger}f_{k,\sigma}+h.c.)\nonumber\\
&+&\frac{Z^{2}}{2}\sum_{k,\sigma}(\Delta \eta_{k}f_{k,\sigma}^{\dagger}f_{-k,-\sigma}^{\dagger}+h.c.)-N\frac{|\Delta|^{2}}{J}\nonumber\\
&+&\lambda\sum_{k,\sigma}(p_{\sigma}^2+p_{-\sigma}^2)-\alpha\sum_{k,\sigma}(p_{\sigma}^2+d^2)\nonumber\\
&+&N\lambda(e^2+d^2-1)+NUd^{2},\
\end{eqnarray}
where $N$ is the number of lattice sites  and $\Delta=\frac{Z^{2}J}{N}\sum_{k}\eta_{k}\langle f_{-k,-\sigma}f_{k,\sigma}\rangle$ represents the superconducting order parameter for extended $s$ or $d$-wave symmetries~\cite{Ptok,Daniel18}, since the strong Coulomb repulsion prevents local $s$-wave pairing between $f$-electrons. $\eta_k$ denotes any of the possible pairing symmetries $\cos k_x+\cos k_y$
and $\cos k_x-\cos k_y$ for $s$ and $d$ waves, respectively.

It is worth to point out that we obtain, for both symmetries, stable superconducting states in different parameter regions, while most of the authors have studied only $d$-wave superconductivity in the presence of strong local repulsive interactions~\cite{Spalek88,Spalek13}. We also have added in the Hamiltonian, Eq.~(\ref{hamilt}), the chemical potential $\mu$ that has to be adjusted when we fix the total band-filling, $n=n_s + n_f$ at different values. We assume a nearest neighbor, constant attractive interaction $J_{ij}=J$. For simplicity, we consider a square lattice with $\epsilon_{k}^{c}=-2t(\cos k_x +\cos k_y )$ and $\epsilon_{k}^{f}=\epsilon_{0}^{f}+\gamma\epsilon_{k}^{c}$, where $\gamma=t_f/t$ ($\gamma<1$) is the ratio of hopping terms of the quasi-particles in different bands and $\epsilon_{0}^{f}$ is the bare $f$-band center. Furthermore, we introduce $\tilde{\epsilon}_{k}^{f}=\epsilon_{k}^{f}+\alpha$ as the renormalized dispersion of the $f$-band. We took the lattice parameter $a=1$.

We use the Green's function method~\cite{Daniel16,Daniel18}, such that the excitations in the superconducting phase are given by the poles of these Green's functions, which in turn are obtained from their equations of motion. These excitations have energy given by $\pm \omega_{1,2}$, where,
\begin{eqnarray}\label{Bogoliubov}
\omega_{1,2}&=&\sqrt{A_k\pm\sqrt{B_k}}\\
A_k&=&\frac{{\varepsilon_{k}^{c}}^{2}+{\varepsilon_{k}^{f}}^{2}}{2}+\tilde{V_k}^{2}+\frac{(\tilde{\Delta}\eta_k)
	^{2}}{2},\nonumber\\
B_k&=&\left(\frac{{\varepsilon_{k}^{c}}^{2}-{\varepsilon_{k}^{f}}^{2}}{2}\right)^2+\tilde{V_k}^{2}
(\varepsilon_{k}^{c}+\varepsilon_{k}^{f})^{2}+\frac{(\tilde{\Delta}\eta_{k})^{4}}{4}\nonumber\\
&-&\frac{(\tilde{\Delta}\eta_k)^{2}}{2}({\varepsilon_{k}^{c}}^{2}-{\varepsilon_{k}^{f}}^{2})+(\tilde{\Delta}\eta_k
\tilde{V_k})^{2},
\end{eqnarray}
with, $\tilde{V_k}=Z V_k$, $\tilde{\Delta}=Z^{2}\Delta$, $\varepsilon_{k}^{c}=\epsilon_{k}^{c}-\mu$,
and $\varepsilon_{k}^{f}=\tilde{\epsilon}_{k}^{f}-\mu$.

The effect of the hybridization's parity on the superconducting properties~\cite{Tremblay,Vojta},  is considered assuming an odd-parity hybridization $\tilde{V}_{-k}=-\tilde{V}_{k}$  with $\tilde{V}_k=i Z V(\sin k_x+\sin k_y)$, and an
even-parity hybridization, such that,  $\tilde{V}_{-k}=\tilde{V}_{k}$  with $\tilde{V}_k=Z V(\cos k_x+\cos k_y)$.
The former case is relevant when the orbitals  involved in the mixing have angular momenta differing by an odd number, as $p$-$d$ or $d$-$f$,  whereas the latter case is considered when the difference in angular momenta is even~\cite{deus}. For completeness, we also consider a constant $k$-independent hybridization. In all cases, $V$ represents the intensity of the hybridization.

Following the slave boson mean-field approximation, the parameters  $e$, $p$, $d$, $\alpha$ and $\lambda$, are  obtained from the minimization of the Hamiltonian, given by Eq.~(\ref{effmodel})~\cite{Daniel16}, with respect to the different slave-boson parameters. This procedure yields a set of coupled equations that has to be solved self-consistently together with the number and gap equations. The last two are given by, 
\begin{eqnarray}
\label{nT}
&n&=1+\frac{1}{N}\sum_k \sum_{\ell=1,2}\frac{(-1)^{\ell}} {2\sqrt{B_k}}\frac{1}{2\omega_{\ell}}\nonumber\\
&\times&\Bigg \{\left(\varepsilon_{k}^{c}+\varepsilon_{k}^{f}\right)\left(\omega_{j}^{2}+\tilde{V_k}^{2}-\varepsilon_{k}^{c}\varepsilon_{k}^{f}\right)-
\tilde{\Delta}^{2}\eta_{k}^{2}\varepsilon_{k}^{c}\Bigg \}\nonumber\\
&\times&\tanh\left( \frac{\beta\omega_{\ell} }{2}\right),\\
\label{gapeqT}
&\frac{1}{J}&=\frac{Z^{4}}{N}\sum_{k}\sum_{\ell=1,2}\frac{\eta_{k}^{2}(-1)^{\ell}} {2\sqrt{B_k}}\left(\frac{\omega_{\ell}^{2}-{\varepsilon_{k}^{c}}^{2} }{2\omega_{\ell}}\right)
\tanh\left( \frac{\beta\omega_{\ell} }{2}\right),\nonumber\\
\end{eqnarray}
respectively. In this equations  $\beta=1/k_B T$, where $k_B$ is the Boltzmann constant and $T$ is the absolute temperature.

The self-consistent numerical solution of the set of coupled equations described above, allows us to obtain the critical superconducting  temperature $T_c$ for different occupations of the bands and types of hybridization. Assuming that the intensity of the latter can be controlled by external pressure, we obtain the pressure dependence of  $T_c$  for different symmetries  and electronic occupations.

\section{Analysis of Results \label{sec3}}

The numerical solution of the self-consistent coupled equations allows us to obtain both the zero and finite temperature phase diagrams of the model for different cases~\cite{Daniel16}. We consider the influence of the parity of the hybridization for both extended $s$-wave and $d$-wave symmetries of the superconducting order parameter,  which are referred as $\Delta_s$ or $s$-wave and $\Delta_d$, respectively.   In all figures below, we take $\epsilon_{0}^{f}=0$, and the ratio of the effective masses $\gamma=0.1$. For this choice of $\epsilon_{0}^{f}=0$ and before turning on the interactions and hybridizations, the half-filled band case corresponds to $\mu=0$. Furthermore, we renormalize all the physical parameters by the $c$-band hopping term  $t=1$.

\subsection{Zero temperature results \label{sec3.1}}
\begin{figure}[h]
	\centering
	\includegraphics[width=1\columnwidth]{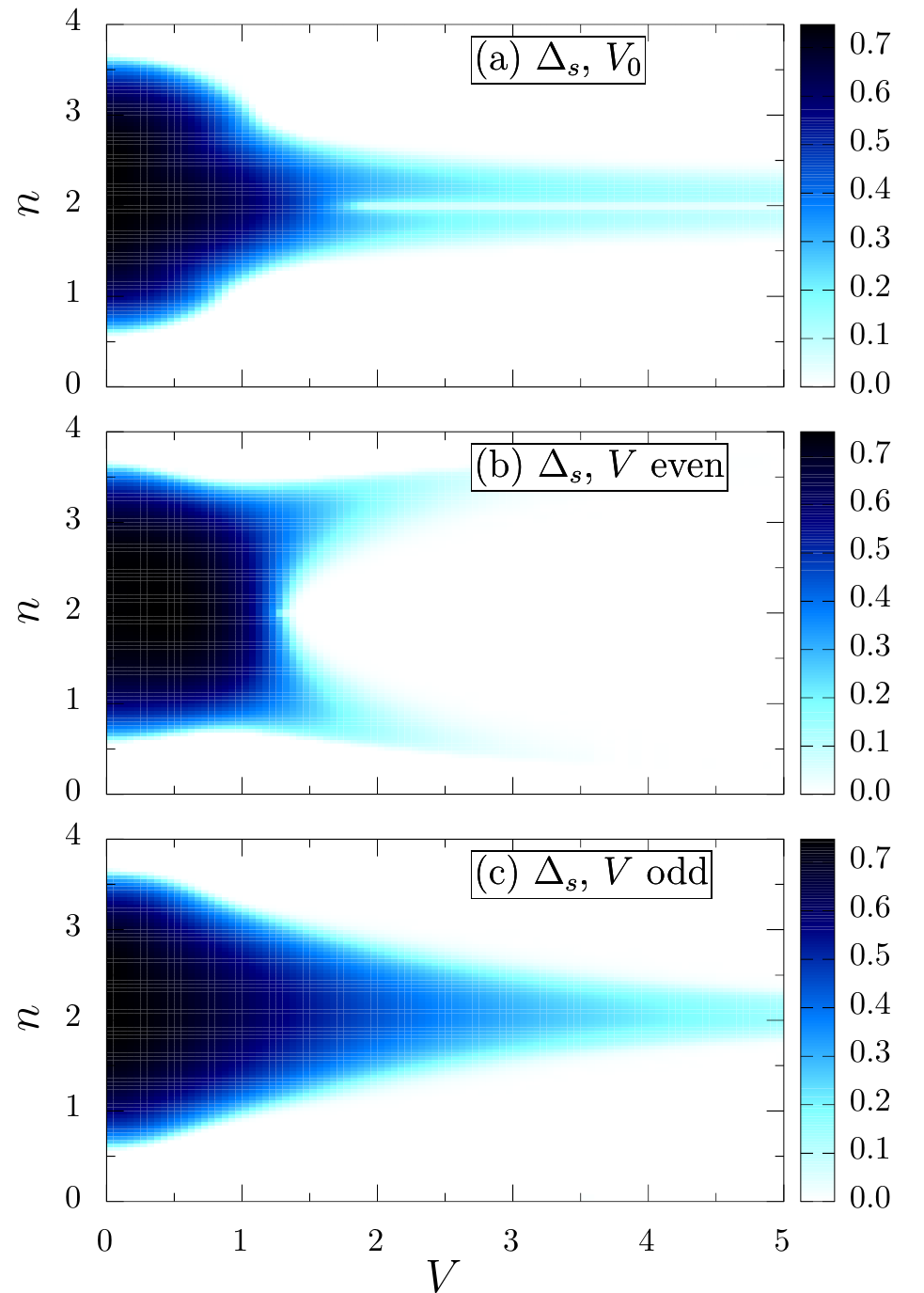}
	\caption{(Color online) Density plots of $\Delta_s$ varying $n$ and $V$,
		for fixed values of attractive interaction $J=-2$ and a Coulomb repulsion $U=1$. Figs.~\ref{fig1}(a), \ref{fig1}(b) and~\ref{fig1}(c) are sketched considering a constant~\cite{Daniel16}, even and odd-parity hybridizations, respectively.}
	\label{fig1}
\end{figure}

The zero temperature density plots for the extended $s$-wave order parameter are shown in Fig.~\ref{fig1} as a function  
of the intensity of hybridization for different band-fillings. Fig.~\ref{fig1}(a), Fig.~\ref{fig1}(b) and Fig.~\ref{fig1}(c) consider the cases of constant ($V=V_0$),  even and odd-parity hybridization, respectively.

\begin{figure}[h]
	\centering
	\includegraphics[width=1\columnwidth]{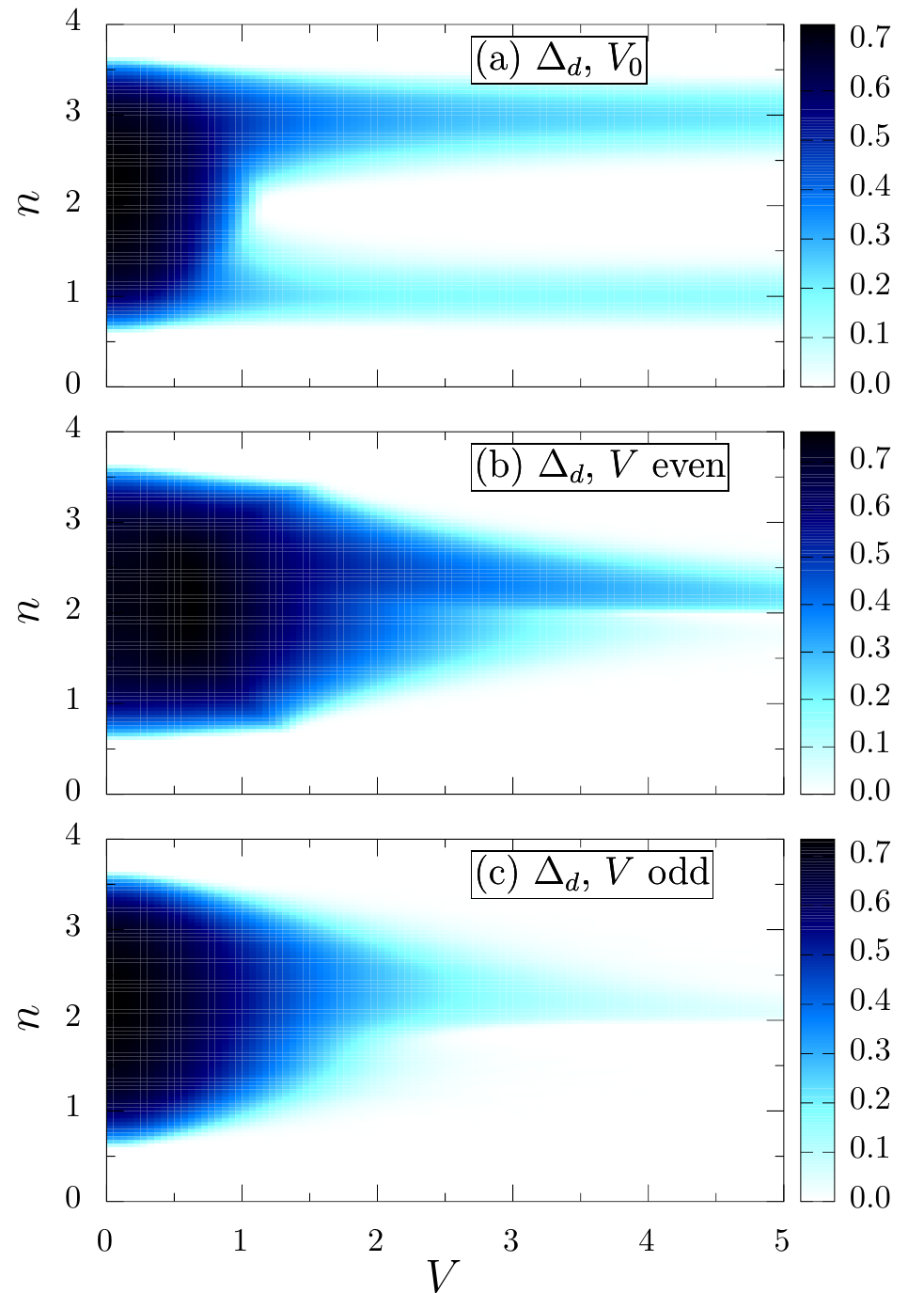}
	\caption{(Color online) Density plots of $\Delta_d$ varying $n$ and $V$,	for fixed values of attractive interaction $J=-2$ and a Coulomb repulsion $U=1$. Figs.~\ref{fig2}(a), \ref{fig2}(b) and~\ref{fig2}(c) are sketched considering a constant~\cite{Daniel16}, even and odd-parity hybridization, respectively.}
	\label{fig2}
\end{figure}

Fig.~\ref{fig2} shows the same density plots, but now for the case of a $d$-wave order parameter, as a function of the intensity of hybridization and  different band-fillings. Fig.~\ref{fig2}(a), Fig.~\ref{fig2}(b) and Fig.~\ref{fig2}(c) consider  constant~\cite{Daniel16}, even and odd-parity hybridizations, respectively.

The influence of the symmetry of the hybridization is notable in the figures, for both cases of $d$ and extended $s$-wave superconductivity. For extended $s$-wave, it is remarkable, in the case of even hybridization, the existence of a critical value of hybridization that suppresses superconductivity. Also the regions of superconductivity are almost symmetric with respect to the half-filling $n=2$, independent of the parity of the hybridization, which is not the case for $d$-wave superconductivity. For $d$-wave and odd hybridization the phase diagram, Fig.~\ref{fig2}(c), is very similar to that of Ref.~\onlinecite{Spalek2} obtained using a full variational Gutzwiller wave function incorporating non-local effects of the on-site interaction. Then, in spite of the very different approaches, the results they yield are in qualitative agreement.

\subsection{Finite temperatures}

Next, we consider the temperature dependence of the self-consistent equations and solve for the superconducting critical temperature ($T_c$) of the model as a function of different parameters. Since $T_c$ is the most accessible experimental quantity and many of the model parameters can be tuned, the calculation of $T_c$ provides a direct test of the results. Besides, since we are studying a non-trivial two-band model, in which superconductivity coexists with strong local correlations and competes with the hybridization between the bands, the study of the effect of each of these features in $T_c$ turns out to be very important. Therefore, in this section we investigate the dependence of the critical temperature $T_c$  on the parity of the hybridization, band-filling and the repulsive interaction for each specific symmetry of the superconducting (SC) order parameter. 
\begin{figure}[!h]
\begin{center}
\includegraphics[width=1\columnwidth]{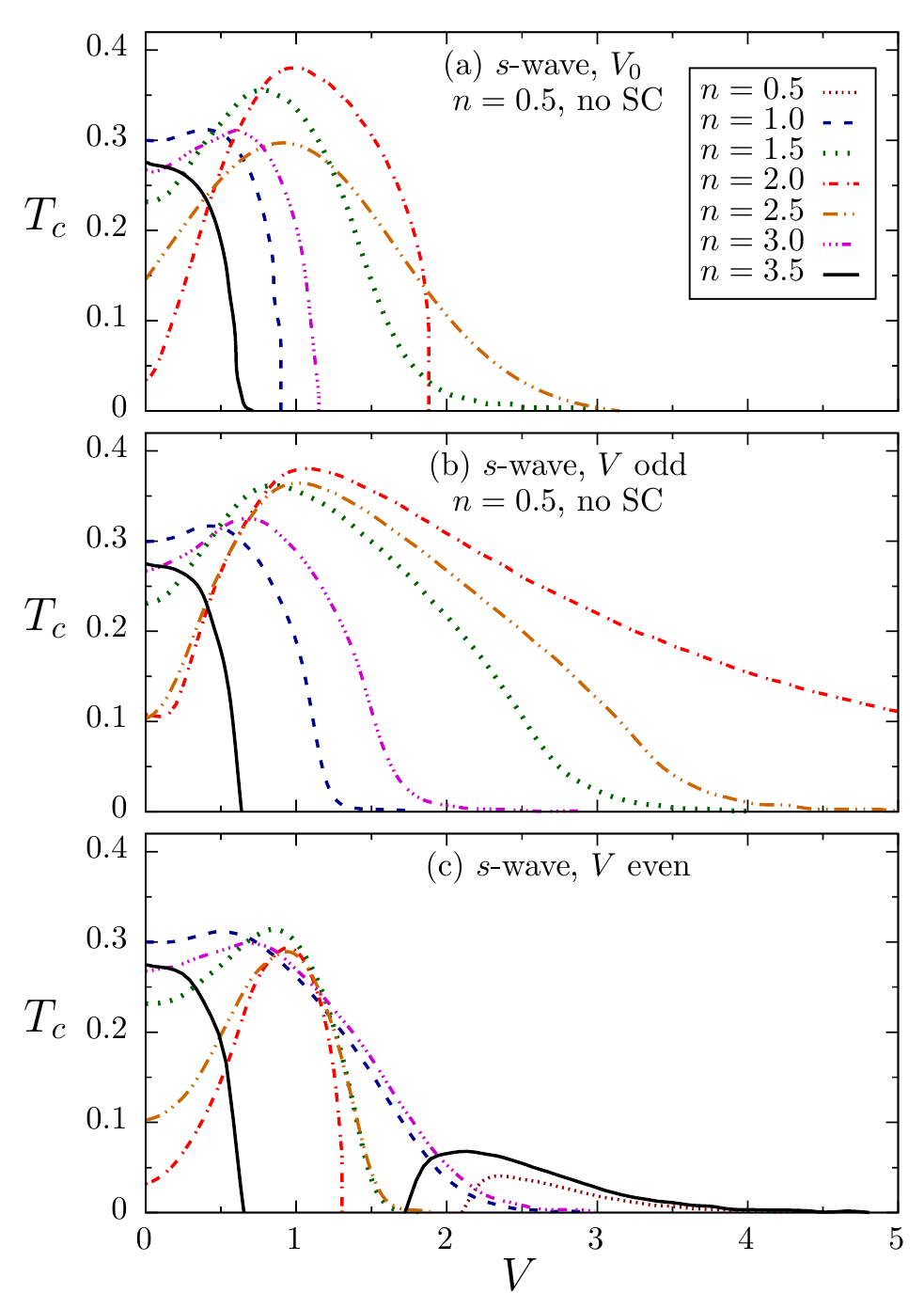} 
\end{center}
\caption{(Color online)  $T_c$ as a function of different $V$ symmetries  considering a extended $s$-wave SC order parameter for fixed values of attractive interaction $J=-2.0$ and Coulomb repulsion $U=1.0$. Figs.~\ref{fig3}(a), \ref{fig3}(b) and~\ref{fig3}(c) are sketched for several values of the band-filling $n$, considering a constant, odd and even-parity hybridization, respectively (see text).}
\label{fig3}
\end{figure}
\subsubsection{The critical temperature as function of the hybridization}
\begin{figure}[!h]
\begin{center}
\includegraphics[width=1\columnwidth]{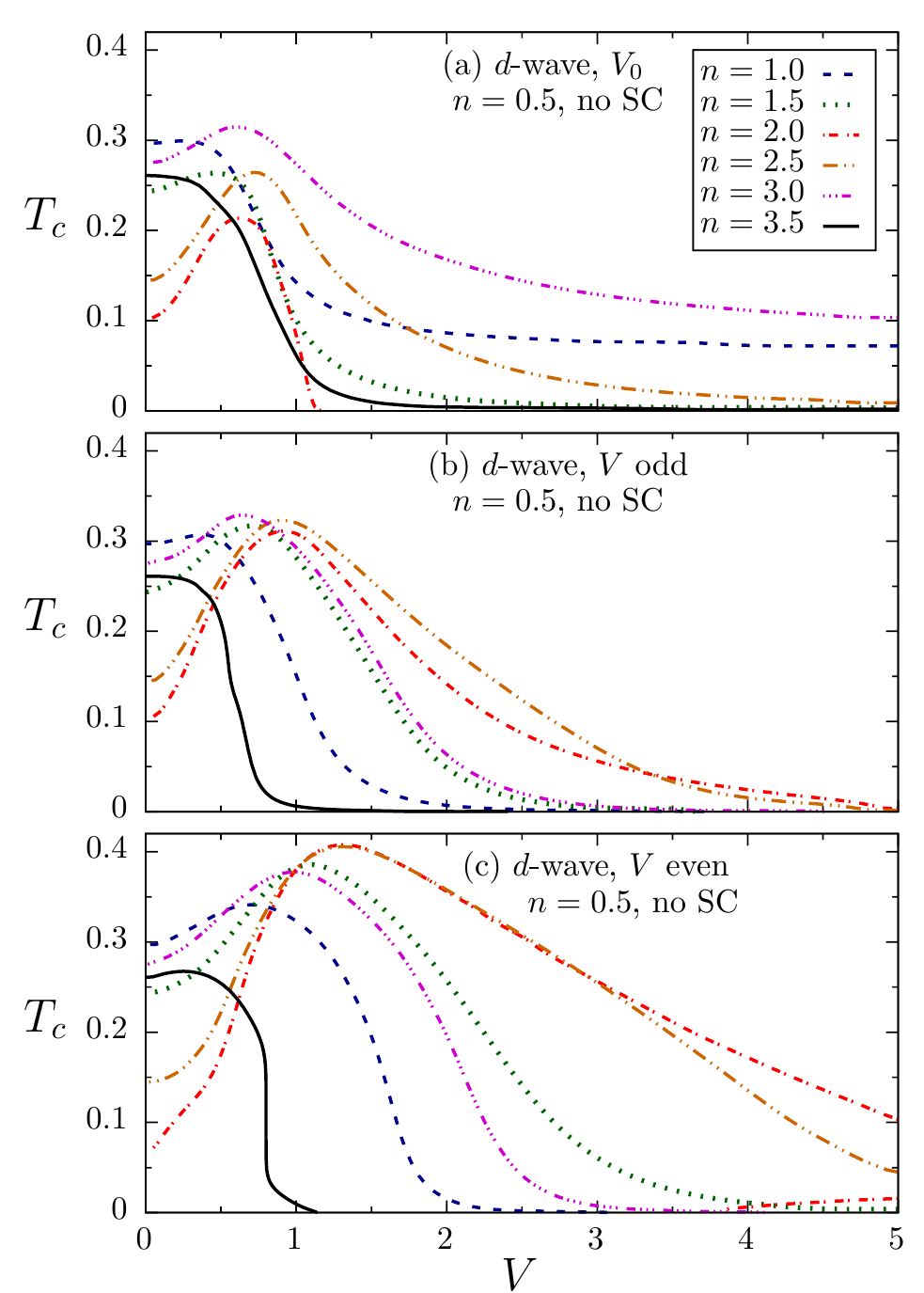} 
\end{center}
\caption{(Color online) $T_c$ as a function of different $V$ symmetries considering a $d$-wave SC order parameter, for fixed values of attractive interaction $J=-2.0$ and Coulomb repulsion $U=1.0$. Figs.~\ref{fig4}(a), \ref{fig4}(b) and~\ref{fig4}(c) are sketched for several values of the band-filling $n$, considering a constant, odd and even-parity hybridization, respectively (see text).}
\label{fig4}
\end{figure}

First, we show the results for $T_c$ as a function of  hybridization for different band-fillings and parities of the mixing. We consider both cases, of extended $s$-wave (Fig.~\ref{fig3}) and $d$-wave (Fig.~\ref{fig4}) symmetries of the  superconducting order parameter.

For consistency, we check our results with those for the case of constant hybridization and extended $s$-wave symmetry (Fig.~\ref{fig3}(a)) that has been calculated previously~\cite{Sacramento}.   The behavior of  $T_c$ that we obtain agrees qualitatively with that in the literature~\cite{Sacramento}. 
In Fig.~\ref{fig3}(b) and Fig.~\ref{fig3}(c), as well as in Fig.~\ref{fig4}(b) and Fig.~\ref{fig4}(c), we show our new results for the critical temperature for different parities of the mixing between the bands and several occupations. Fig.~\ref{fig3} shows the behavior of the critical temperature for a $s$-wave superconducting order parameter as a function of the parity of the hybridization and  different band-fillings. Fig.~\ref{fig4} shows the same results for $T_c$, but now for a $d$-wave superconducting order parameter.  It is interesting that even for a rather \textit{dilute} case $n=0.5$, superconductivity arises for large values of an even hybridization ($V>2$), see Fig.~\ref{fig3}(c).
For larger band-fillings, the superconducting phase appears for all $V$ symmetries and remains for occupations up to  $n \approx 3.5$.

One can see from Fig.~\ref{fig3} and  Fig.~\ref{fig4} that, except for $n \approx 0.5$ and $n \approx 3.5$, there is a region in the phase diagrams where the critical temperatures increase with increasing the intensity of hybridization from small V. This type of behavior has been seen in previous works~\cite{Sacramento}. In order to understand the physics behind this increase, we have studied the behavior of the free energy~\cite{Robas} as a function of the superconducting order parameter in the regions of the phase diagrams where this enhancement of $T_c$ is observed.  

\begin{figure}[!h]
\begin{center}
\includegraphics[width=1\columnwidth]{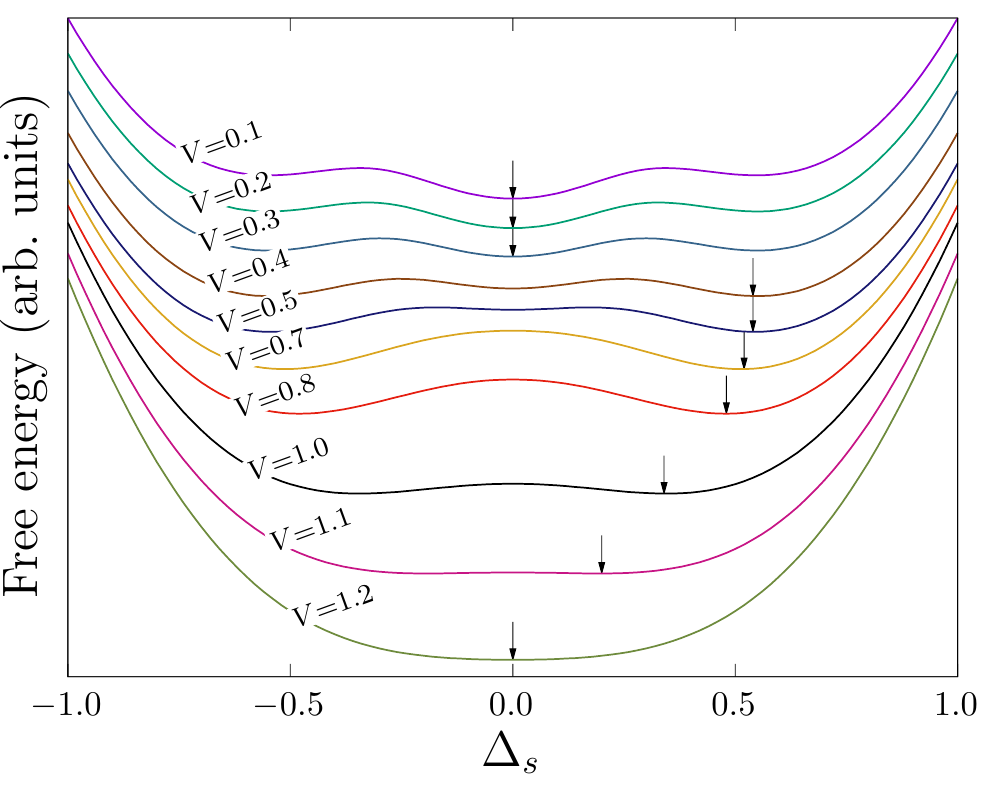} 
\end{center}
\caption{(Color online) Free energy variation for different values of $\Delta_s$. The minimum (arrow) can be seen floating abruptly and then smoothly when we fix $T=0.25$, $n=1.5$, and we vary $V$ (see text).}
\label{fig5}
\end{figure}

Fig.~\ref{fig5} shows the free energy as a function of the extended $s$-wave order parameter $\Delta_s$, for a fixed temperature $T=0.25$ and $n=1.5$, varying the intensity of the even hybridization from $V=0.1$ to $V=1.2$ (see also Fig.~\ref{fig3}(c)). One can see the presence of minima (arrow), located at zero and finite $\Delta_s$, that exchanges stability as the intensity of the  even parity hybridization increases.  For small values of $V$, the stable phase is the normal metal with $\Delta_s=0$, although one can already notice the presence of two metastable minima for finite $\Delta_s$.  As $V$ reaches $V \approx 0.4$ the three minima become degenerate and the system enters in the superconducting phase through a first-order transition. 

First-order transitions between normal and superconducting states have already been obtained in strongly correlated systems~\cite{Vojta} using a Kondo lattice approach. Here, we obtain coexistence between these states in a two-band model. Experimentally, first-order transitions have  been reported in compounds that present spin-triplet superconductivity~\cite{Yonezawa,Kittaka} and in pnictides, where  structural and magnetic first-order phase transitions occur for a characteristic temperature~\cite{Li,Wang}. 

\begin{figure}[!h]
\begin{center}
\includegraphics[width=1\columnwidth]{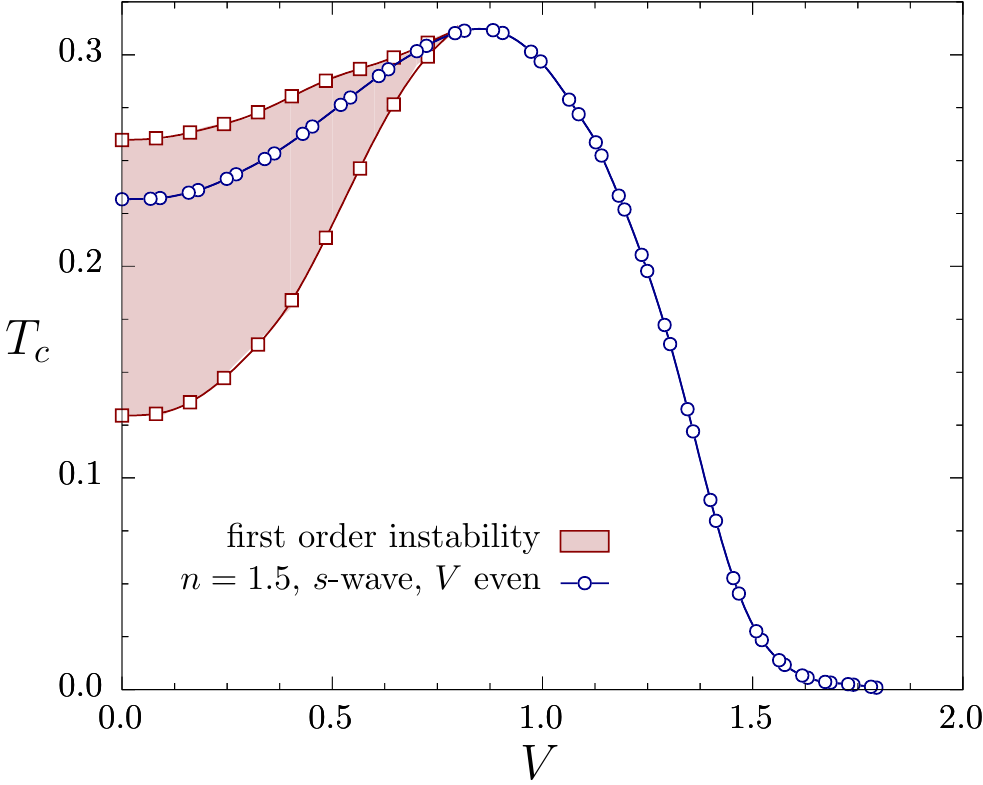} 
\end{center}
\caption{(Color online) Phase diagram  showing $T_c$ as a function of even parity hybridization for extended $s$-wave SC order parameter. The curves are obtained from  plots  of the free energy as in Fig.~\ref{fig5}. The shaded area represents the region of phase coexistence.}
\label{fig6}
\end{figure}

For $V \approx 0.4$ there is an exchange of stability between the normal and superconducting phases.  Further increase of the value of $V$ takes the system  smoothly, through a second-order transition,  to the normal phase. In general, we can conclude that in the regions of the phase diagram that $T_c$ increases with $V$ the system is metastable and presents a first-order phase transition and coexistence of phases. 
The actual phase diagram obtained from an analysis of the free energy as a function of the superconducting order parameter is shown in Fig.~\ref{fig6}. The shaded region represents the place of coexistence.
This kind of behavior, i.e., first-order transitions for small $V$, up to the maxima in the $T_c$ versus $V$ curves will be present for all symmetries. However, another notable feature worth to emphasize is that the first-order region, associated with the existence of maxima in $T_c$ versus $V$ curve, is suppressed for small and large band-fillings. In fact for $n \lesssim 0.5$ and $n \gtrsim 3.5$, we no longer observe maxima in $T_c$ as a function of $V$ for any symmetry (see Fig.~\ref{fig3} and Fig.~\ref{fig4}), and consequently we have only second-order transitions. Our numerical results show that the local Coulomb repulsion decreases the region in the phase diagram where the system presents first-order transitions, but does not suppress it. In practice, we observe that the maxima in the  $T_c$ versus  $V$ curves shifts to smaller values of $V$ with increasing Coulomb repulsion. It is worth noticing in Fig.~\ref{fig3}(c), for $V$ even and large band-filling ($n = 3.5$), the presence of two superconducting  \textit{domes} as a function of $V$, unlike for constant (Fig.~\ref{fig3}(a)) or odd (Fig.~\ref{fig3}(b)) parity of the hybridization.

Finally, we remark that for $d$-wave symmetry, Fig.~\ref{fig4}, we observe a different behavior from that of the extended $s$-wave case. First, for $n=0.5$, there is no SC for any parity of the hybridization and second we find no evidence of two superconducting domes for  $V$ even and large occupations, as in the previous case of extended $s$-wave superconductivity. On the other hand, the first-order transitions remain present for small $V$ associated with a region of the phase diagram where $T_c$ increases with $V$, the intensity of hybridization. Once again, this instability region is suppressed as $n$ increases. The effect of increasing the local $U$ is similar to the extended $s$-wave case and independent of the parity of $V$, i.e., the instability region decreases, but is not suppressed.
\begin{figure}[!b]
\begin{center}
\includegraphics[width=1\columnwidth]{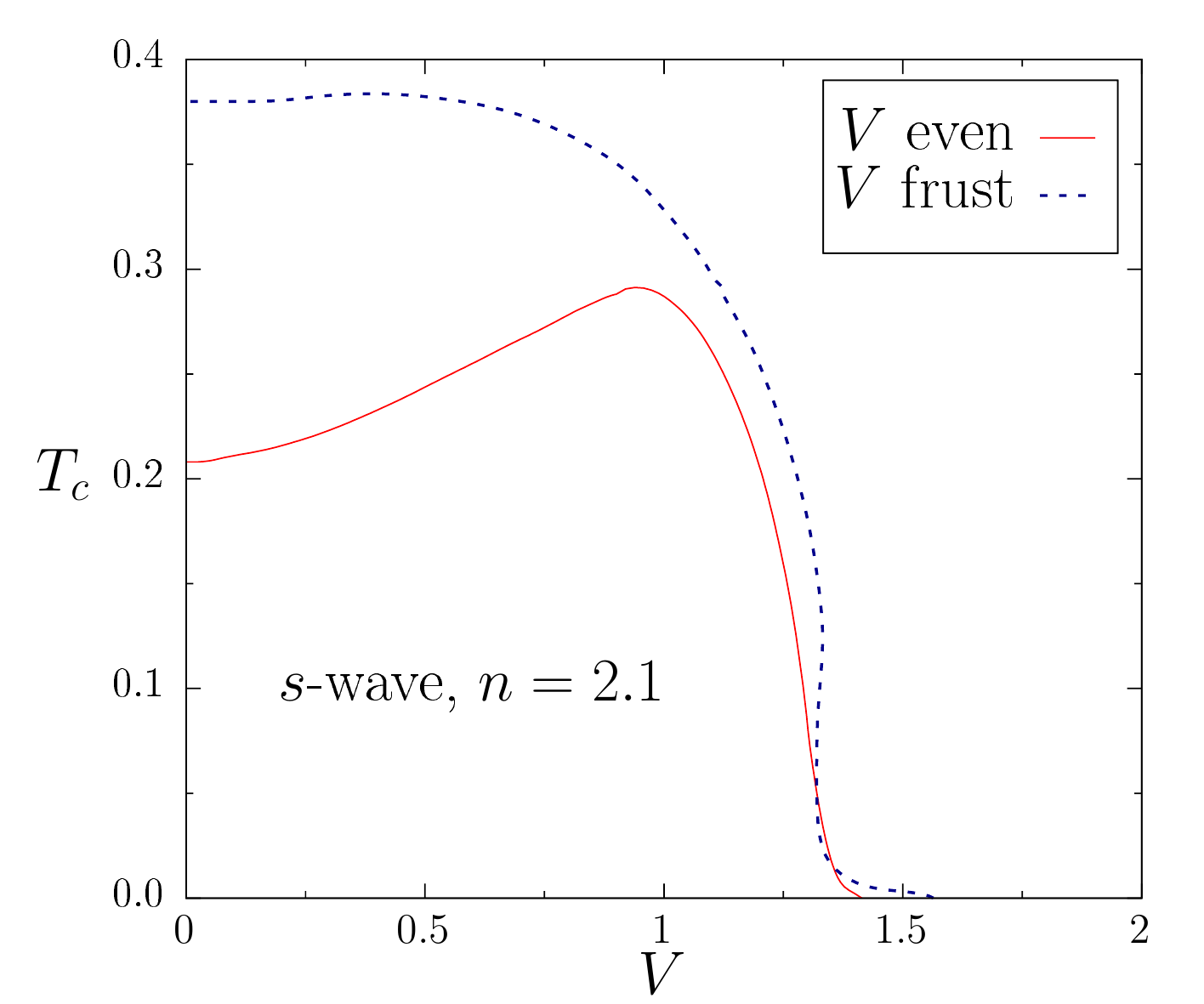} 
\end{center}
\caption{(Color online) $T_c$ as a function of hybridization for the cases of no frustration (continuous line) and increasing frustration (dashed line) for a band-filling of $n=2.1$ and extended $s$-wave pairing and  fixed values of attractive interaction $J=-2$ and Coulomb repulsion $U=1$ (see text).}
\label{fig7}
\end{figure}

The model investigated here also supports antiferromagnetic solutions for similar values of the parameters~\cite{Sacramento1}. This magnetic phase competes with superconductivity in the same region of the phase diagram. In multi-band systems as heavy fermions, superconductivity arises in proximity to an antiferromagnetic quantum critical point~\cite{matur}. In order to verify whether these two phases actually compete or antiferromagnetic fluctuations enhance superconductivity, we introduced frustration in our model~\cite{Tremblay}. This is done by including a constant hybridization $V_0=1$ and a nearest neighbor symmetric mixing $V(\cos k_x + \cos k_y)$ of varying intensity (see Fig.~\ref{fig7}).

We find that increasing frustration enhances and stabilizes superconductivity, at least for small $V$, as expected if this is in competition with antiferromagnetism. Fig.~\ref{fig7} shows $T_c$  for a band-filling $n=2.1$, in the cases of a pure even hybridization (no frustration) (continuous line) of intensity $V$ and when this competes with a constant one of unit intensity, $V_0=1$ (dashed line). In the former case the transition to the superconducting state is first order below the maximum, as shown by an analysis of the free energy (see Fig.~\ref{fig6}). As frustration is included, $T_c$ increases and the transition becomes second order, such that, frustration stabilizes the superconducting state~\cite{Tremblay}.

\subsubsection{The critical temperature as function of the band-filling}

Next, we show our results for the critical temperature as a function of band-filling, for different intensities and parities of the hybridization and distinct symmetries of the order parameter. Our results agree  qualitatively, when available, with those obtained previously~\cite{Sacramento}.
\begin{figure}[!h]
\begin{center}
\includegraphics[width=1\columnwidth]{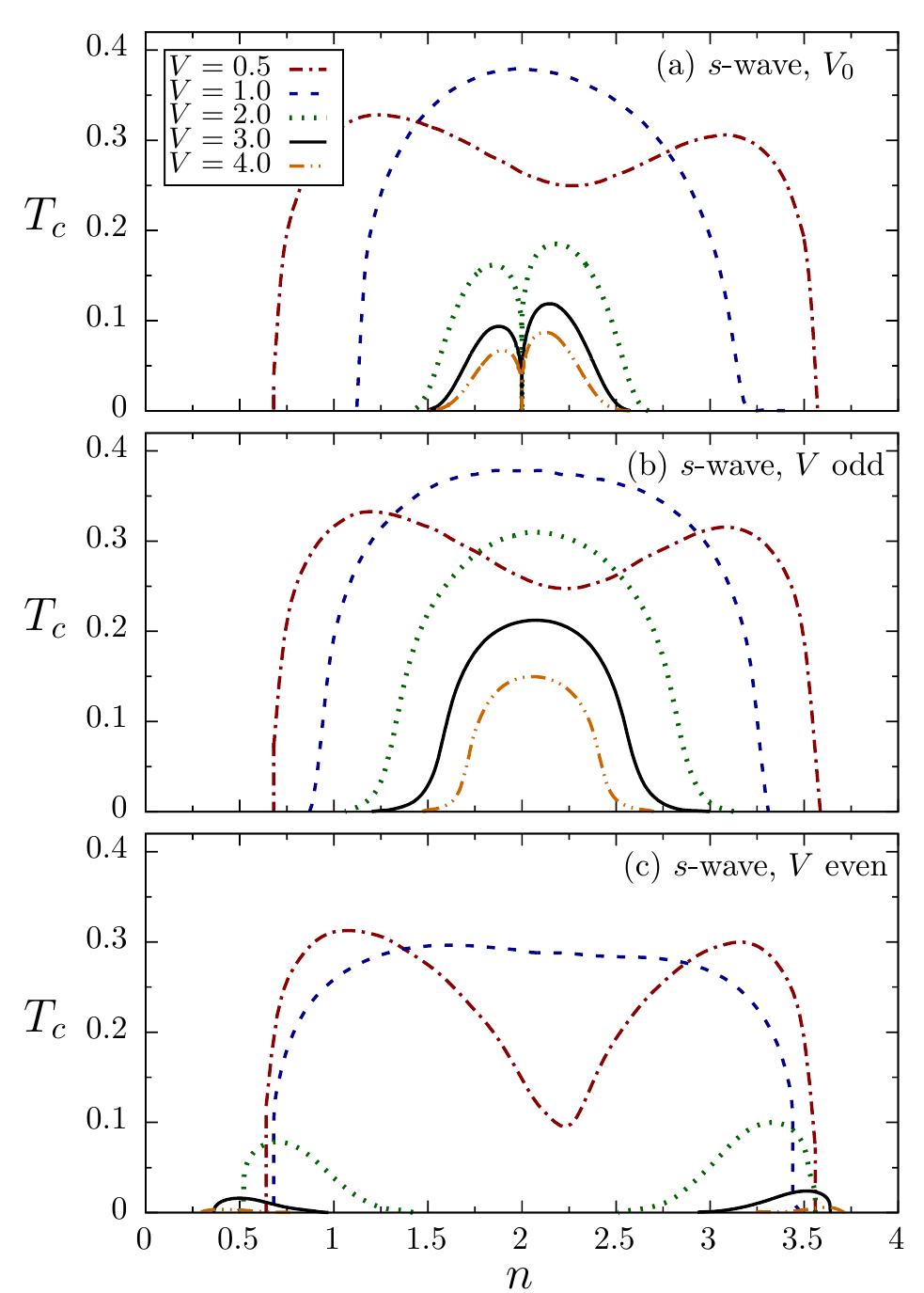} 
\end{center}
\caption{(Color online) $T_c$ as a function of band-filling $n$ considering a extended $s$-wave SC order parameter. This plot is showed for fixed values of attractive interaction $J=-2$ and Coulomb repulsion $U=1$. Figs.~\ref{fig8}(a), \ref{fig8}(b) and~\ref{fig8}(c) are sketched for several values of intensity of $V$, considering a constant, odd and even-parity hybridization, respectively (see text).}
\label{fig8}
\end{figure}
In Fig.~\ref{fig8} and Fig.~\ref{fig9}, we show the finite temperature phase diagrams for different symmetries of the order parameter as a function of band-filling for different parities of the hybridization. The Coulomb repulsion is kept fixed at $U=1$. The phase diagrams for each symmetry depend  strongly on the parity of $\tilde{V_k}$. 
\begin{figure}[!h]
\begin{center}
\includegraphics[width=1\columnwidth]{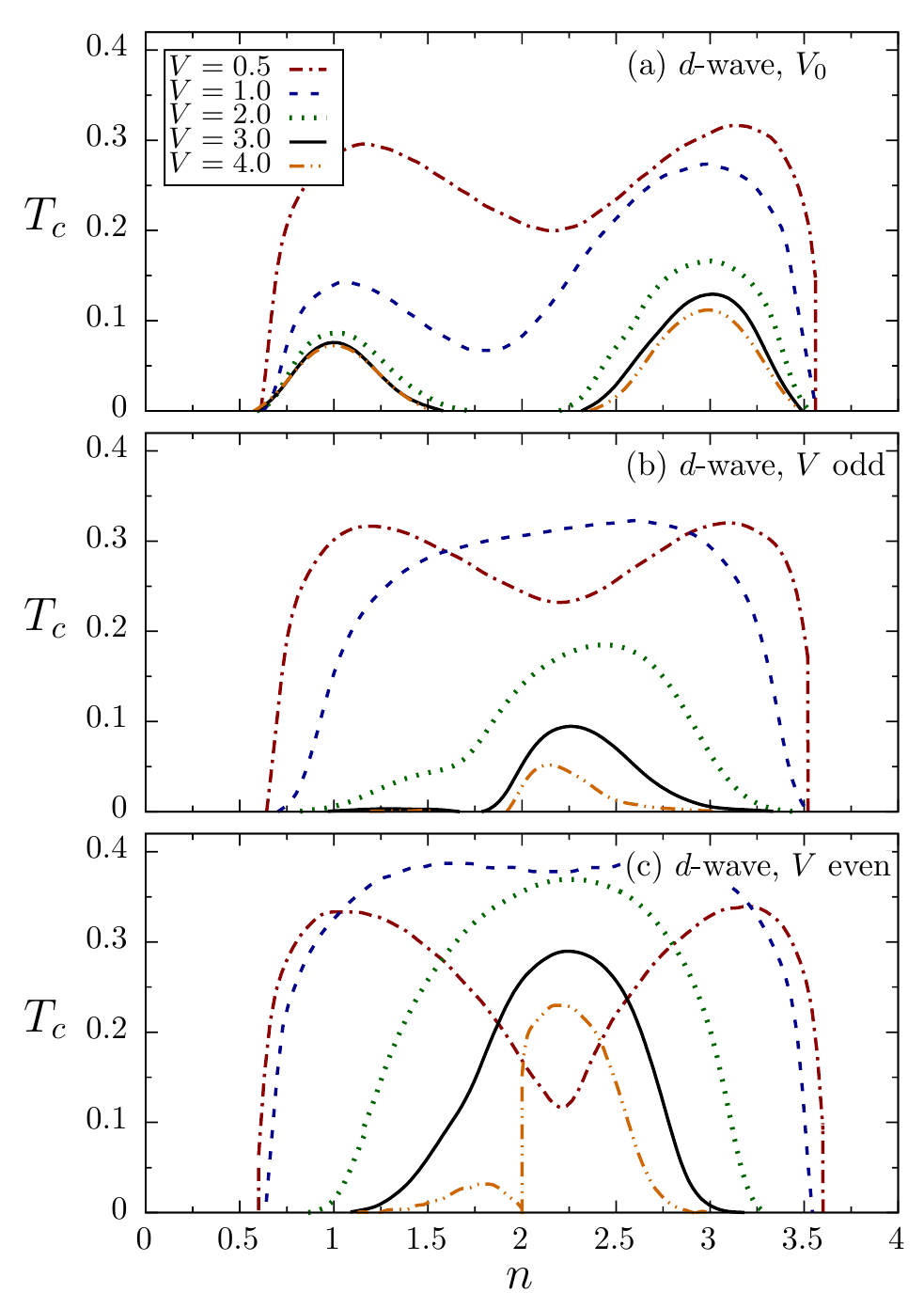} 
\end{center}
\caption{(Color online) $T_c$ as a function of band-filling $n$ considering a $d$-wave SC order parameter. This plot is showed for fixed values of attractive interaction $J=-2$ and Coulomb repulsion $U=1$. Figs.~\ref{fig9}(a), \ref{fig9}(b) and~\ref{fig9}(c) are sketched for several values of intensity of $V$, considering a constant, odd and even-parity hybridization, respectively (see text). }
\label{fig9}
\end{figure}
In general, superconductivity does not occur for small number of particles or holes. In some cases the critical temperature is a maximum for half-filled bands, while for others it is suppressed in this case.  

An analysis of the free energy curves as a function of the band-filling $n$ shows that for small values of the intensity of the hybridization, $V \le 1$, independent of its parity  and symmetry of the superconducting order parameter,  superconductivity becomes metastable for $1.0 \le n \le 3.0$. For $V \ge 1$ coexistence of phases disappears and  superconductivity is stable when $T_c$ is finite.

The results of Fig.~\ref{fig8} for extended $s$-wave symmetry of the order parameter are closely related to those of Fig.~\ref{fig1}. In particular  we remark that for odd parity hybridization, i.e., Fig.~\ref{fig8}(b), the critical temperature attains a maximum for band-filling $n=2$, while for $V$ even it is mostly suppressed for this occupation, even giving rise to two separate regions of superconductivity, see Fig.~\ref{fig8}(c). 
It is a general feature in Fig.~\ref{fig8} that increasing the intensity of hybridization, shrinks the region of superconductivity in the phase diagram. 

The results for $d$-wave symmetry of the order parameter are shown in Fig.~\ref{fig9}, and are closely associated to those of  Fig.~\ref{fig2}. In this case, we notice that the phase diagram is in general asymmetric with respect to half-occupation of the bands ($n=2$), specially for large $V$.  For $n < 2$ and in  both cases of even and odd parity, large intensities of the hybridization suppress superconductivity  as can be seen in  Fig.~\ref{fig9}(c) and Fig.~\ref{fig9}(b), respectively. The values of $T_c$ in this case are significantly higher for even parity hybridization.  For large, $k$-independent  $V$, we see again the presence of two superconducting domes. 

\subsubsection{The critical temperature as function of the Coulomb correlation}

\begin{figure}[!h]
\begin{center}
\includegraphics[width=1\columnwidth]{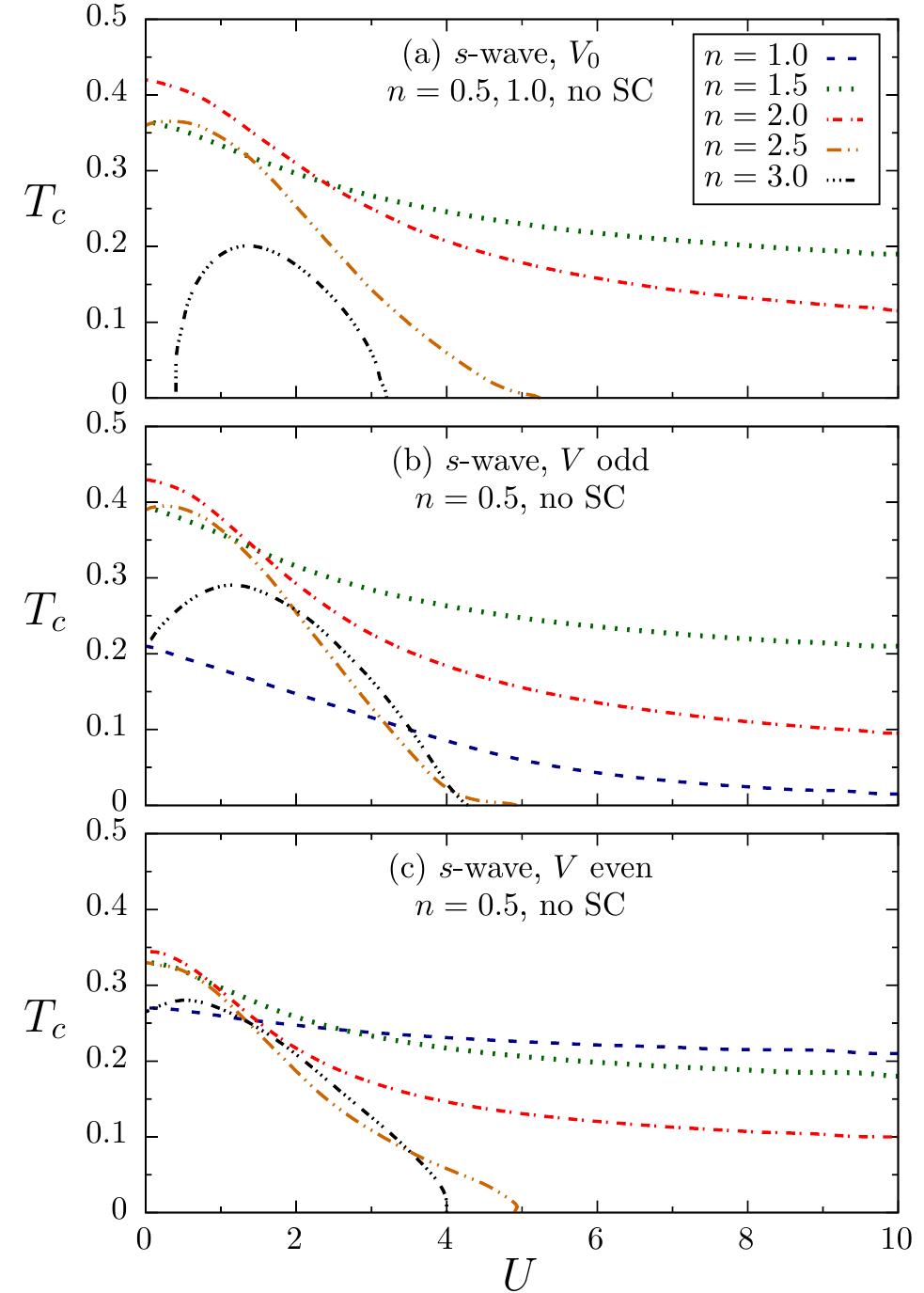} 
\end{center}
\caption{(Color online) $T_c$ as a function of $U$ considering a extended $s$-wave SC order parameter. This plot is showed for fixed values of attractive interaction $J=-2$ and $V=1.0$. Figs.~\ref{fig10}(a), \ref{fig10}(b) and~\ref{fig10}(c) are sketched for several values of $n$, considering a constant, odd and even-parity hybridization, respectively (see text).}
\label{fig10}
\end{figure}

Finally, we discuss the effect of the local Coulomb correlation $U$ on superconductivity.
We start with the case of an extended $s$-wave superconductor where the critical temperatures as a function of $U$ are shown in Fig.~\ref{fig10} for fixed intensity of hybridization ($V=1$). In general, the critical temperature decreases smoothly with increasing local Coulomb repulsion. Remarkably, this is not always the case. For large band-fillings, the critical temperature may increase with $U$  and then be strongly suppressed with further increase giving rise to a superconducting quantum critical point.

Our numerical results for constant $k$-independent hybridization are in qualitative agreement with those obtained previously~\cite{Sacramento} for extended $s$-wave symmetry and large $U$. Here we extend the calculations for a larger range of band-fillings, for different parities of the hybridization and different symmetries of the superconducting order parameter. Also we consider the case of small $U$ that has not been studied before. This is probably the most interesting, since it can give rise to an increase of $T_c$, for sufficiently large occupations.

\begin{figure}[!h]
\begin{center}
\includegraphics[width=1\columnwidth]{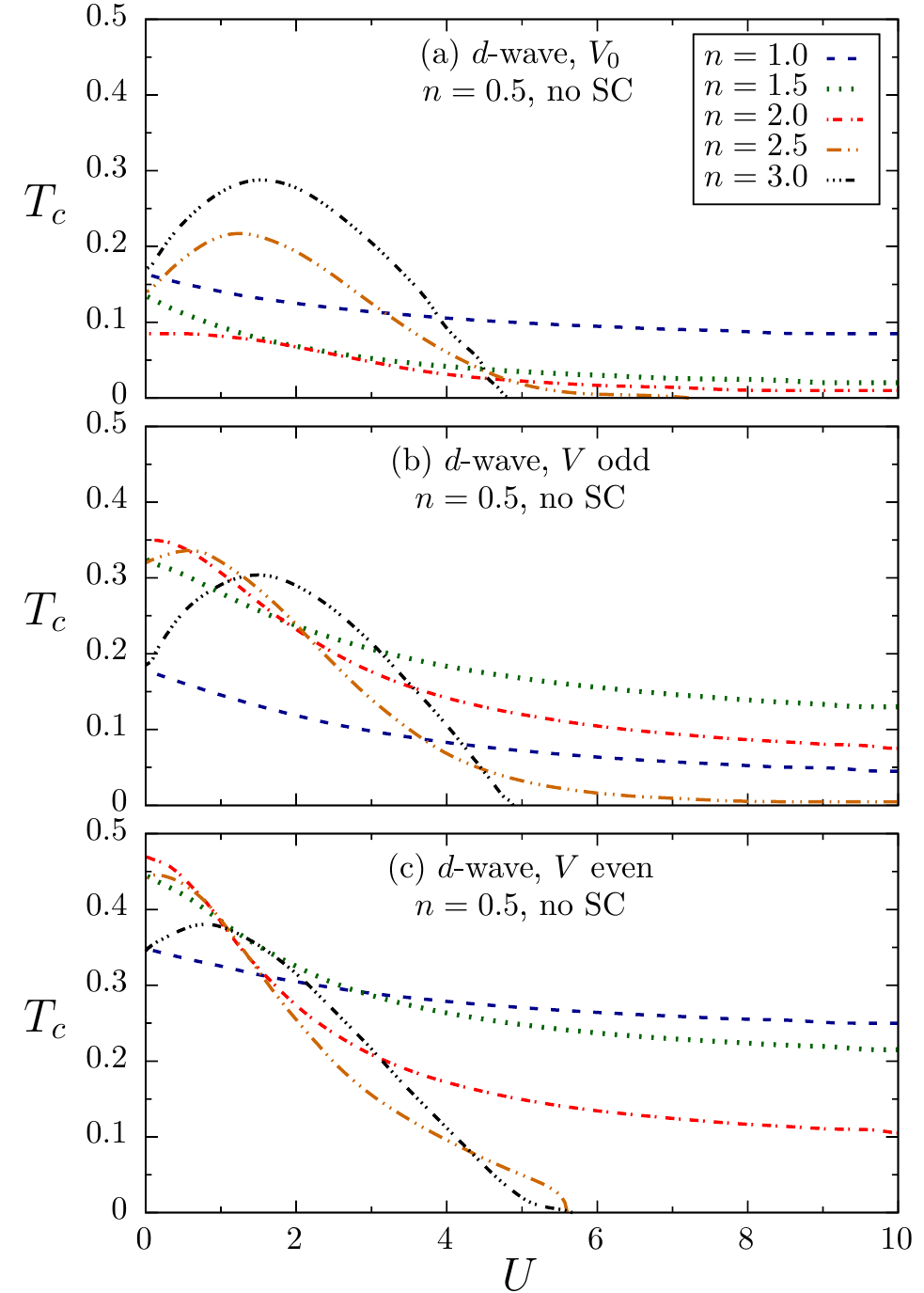} 
\end{center}
\caption{(Color online) $T_c$ as a function of $U$ considering a $d$-wave SC order parameter. This plot is showed for fixed values of attractive interaction $J=-2$ and $V=1.0$. Figs.~\ref{fig11}(a), \ref{fig11}(b) and~\ref{fig11}(c) are sketched for several values of $n$, considering a constant, odd and even-parity hybridization, respectively (see text). }
\label{fig11}
\end{figure}

We further remark in Fig.~\ref{fig10}, that for small band-filling $n=0.5$ there is no SC order and also for $n=1.0$ with $V$ constant, see Fig.~\ref{fig10}(a). As we increase $n$, the finite critical temperature decreases with increasing $U$, but remains finite for large $U$  for specific values of band-filling ($n \leq 2.0$). However, for large band-filling $n=3.0$, we see a very different behavior, i.e., $T_c$ increases and then is suppressed with increasing $U$.  Note that $T_c$ for even parity $V$, Fig.~\ref{fig10}(c), is rather lower than for the other cases.

For $d$-wave symmetry  Fig.~\ref{fig11} shows  results not very different from the extended $s$-wave case. For small band-filling $n=0.5$ there is no SC order for any parity of $V$. As we increase $n$ the SC order appears and $T_c$ decreases, but remains finite for large $U$ only for specific values of band-filling. In contrast with extended $s$-wave symmetry, now the higher $T_c$ is obtained for $V$ even. Here, we also observe an increase of  $T_c$ with band-filling, but there is no longer any $T_c$ \textit{dome} for the same range of parameters. Finally, we point out that a stability analysis of the free energy shows that  superconductivity, depicted in Fig.~\ref{fig10} and Fig~\ref{fig11} as a function of increasing local repulsion and for $V \ge 1$ is always stable. The phase transitions shown in these figures are continuous, 
second-order transitions.

\section{Conclusions \label{sec4}} 

In this work, we have studied a two-band model for a superconductor in a square lattice. Electronic correlations in the narrow band were treated using a slave boson approach. We have obtained  the superconducting order parameter and the critical temperature of the model as a function of the band-filling, hybridization and Coulomb repulsion for both extended $s$-wave and $d$-wave symmetries. We considered the cases of even and odd hybridizations with respect to inversion symmetry.   A superconducting phase is found for all cases, and in some of them, a superconducting quantum critical point is obtained.

For band-fillings between $n\approx[1.0 : 3.0]$, the critical temperature increases as hybridization increases, reaching a maximum and then  decreases.   This initial increase for small  $V$  is generally associated with a metastable character of the superconducting state and the presence of first-order transitions in this region of the phase diagram.
When frustration effects are included, the critical temperature no longer increases and the superconducting phase is \textit{stabilized}. 

We found that the critical temperature is strongly dependent on the band-filling. For instance, for $n\approx 0.5$, as well as $n\approx3.5$ and even symmetry of hybridization, two  superconducting regions are obtained: one with the usual initial decrease of the superconducting critical temperature  with $V$ and a \textit{second} with a dome for higher values of  hybridization. The Coulomb repulsion is also an important ingredient affecting superconductivity: for certain values of band-filling it can lead to a suppression of the superconducting phase, instead of a continuous asymptotic decrease as reported in previous works~\cite{Daniel16,Sacramento}.

We expect that our results can be useful as a guide to expectations for the finite temperature properties of multi-band superconductors. The behavior of the critical temperature with external parameters that can be controlled experimentally, together with  the theoretical insights that we have obtained can provide a useful criterion to distinguish between different symmetries of the order parameter and the nature of the orbitals involved in the superconductivity.

\section{ACKNOWLEDGMENTS}

M.A.C. would like to thank the Brazilian agencies FAPERJ, CAPES and CNPq for partial financial support. N.L. would like to thank to CNPq for doctoral fellowship. D.R. would like to thank the Brazilian Center for Research in Physics (CBPF) where part of this work was done. Finally, we would like to thank COTEC (CBPF), since the numerical calculations were performed on the \textit{Cluster HPC}.

\end{document}